\documentclass[twocolumn,showpacs,preprintnumbers,amsmath,amssymb,a4paper,floatfix]{revtex4}
\usepackage[dvips]{graphicx}
\usepackage{dcolumn}
\usepackage{bm}

\newcommand{\ket}[1]{\left | \, #1 \right \rangle}
\newcommand{\bra}[1]{\left \langle #1 \, \right |}

\begin{document}

\author{Jakub S. Prauzner-Bechcicki}
\affiliation{Instytut Fizyki imienia Mariana Smoluchowskiego,
  Uniwersytet Jagiello\'nski,\\
 ulica Reymonta 4, PL-30-059 Krak\'ow, Poland}

\title{Two-mode squeezed vacuum state coupled to the common thermal reservoir}

\date{\today}

\begin{abstract}
 Entangled states play a crucial role in quantum information protocols, thus the dynamical behavior of
entanglement is of a great importance. In this paper we consider a two-mode squeezed vacuum state coupled
to one thermal reservoir as a model of an entangled state embedded in an environment. As a criterion for
entanglement we use a continuous-variable equivalent of the Peres-Horodecki criterion, namely the Simon
criterion. To quantify entanglement we use the logarithmic negativity. We derive a condition, which assures that the state remains entangled in spite of
the interaction with the reservoir. Moreover for the case of interaction with vacuum as an
environment we show that a state of interest after infinitely long interaction is not only entangled,
but also pure. For comparison we also consider a model in which each of both modes
is coupled to its own reservoir. 
\end{abstract}
\pacs{03.65.Ud, 03.67.-a, 03.67.Mn, 42.50.Dv}

\maketitle

\section{Introduction}
A number of possible practical applications of entangled states has
been proposed, including the quantum computation~\cite{Bar} and the quantum teleportation~\cite{Ben,braun}.
 Thus it is very important to know, how the interaction with
 an environment influences the dynamical behavior of entanglement. In many experiments as a source
  of entangled pairs serves the process of the parametric down conversion in the non-linear
 crystals~\cite{wu,Fur}, in which two-mode squeezed vacuum states are produced. Hence, here we 
 consider the time evolution of the two-mode squeezed vacuum, as an example of continuous
 variable entangled state, embedded in a thermal environment. Moreover, let us assume that, for some
 reason, e.g. storage, both modes are captured in a resonant cavity, i.e. they are coupled to
  a common reservoir. Additionally we presume that they have the same frequency.
 There have been intensive research in this direction, but
 Jeong \textit{et al.}~\cite{kim2}, Kim and Lee~\cite{kim}, Scheel \textit{et
al.}~\cite{welsch}, Hiroshima~\cite{hiroshima} and Paris~\cite{paris} assumed that the two-mode squeezed vacuum
 interacts with the environment in a different way. In their model both modes are spatially separated and thus each
  of them is independently coupled to
its own reservoir. Both modes are taken to have the same frequency and the same coupling
constant. This model is justified, while the two-mode squeezed vacuum can be used as a quantum channel
between sender and receiver stations in the quantum teleportation~\cite{braun}. There is a possible experimental investigation of degradation of
 entanglement due to the interaction with the environment~\cite{Bowen}.

\section{Model and description}
In this paper we consider a bipartite system, namely, two modes of the electromagnetic field. Each of those modes has the
Hilbert space equivalent to the Hilbert space of the harmonic oscillator. The Hilbert space of the system of the interest is
a tensor product of subsystem's Hilbert spaces, $\mathcal{H}=\mathcal{H}_1\otimes\mathcal{H}_2$. Each mode is described by
dimensionless quadrature operators defined as $\hat{x}_j = (\hat{a}_j+\hat{a}_j^{\dagger})/ \sqrt{2}$
and $\hat{p}_j = i (\hat{a}_j^{\dagger}-\hat{a}_j)/ \sqrt{2}$ ($j=1, 2$). Those operators obey
a standard commutation relation $[\hat{x}_j, \hat{p}_k]=i\delta_{jk}$, which is equivalent to
$[\hat{a}_j,\hat{a}_k^{\dagger}]=\delta_{jk}$.

Bipartite system in question is embedded in the thermal environment, which we imitate by the set of
harmonic oscillators. We model the coupling of both modes to the common
reservoir by the interaction Hamiltonian of the form:
\begin{equation}\label{jedenham}
H^{int}_{one}=\eta\sum_{i=1,2}(\hat{a}_i\sum_{k=1}^{\infty}\hat{b}_k^{\dagger}+\hat{a}_i^{\dagger}\sum_{k=1}^{\infty}\hat{b}_k).
\end{equation}
While in the case when each mode interacts with its own reservoir, the interaction Hamiltonian
is:
\begin{equation}\label{dwaham}
H^{int}_{two}=\eta(\hat{a}_1\sum_{k=1}^{\infty}\hat{b}^{\dagger}_k+\hat{a}^{\dagger}_1\sum_{k=1}^{\infty}\hat{b}_k+\hat{a}_2\sum_{l=1}^{\infty}\hat{c}^{\dagger}_l+\hat{a}^{\dagger}_2\sum_{l=1}^{\infty}\hat{c}_l)
\end{equation}
In both cases $\eta$ is a coupling constant.

In order to study the evolution of our system we use the Wigner distribution function defined
as~\cite{Wigner}:
\begin{eqnarray}
W(x,p)=\pi^{-2} \int d^2x' \bra{x-x'}\hat{\rho}\ket{x+x'}\nonumber\\
\times \exp(2ix'\cdot p),
\end{eqnarray} 
where $\hat{\rho}$ is density operator, $x=(x_1,x_2)$ and \mbox{$p=(p_1,p_2)$}.

For the sake of convenience we arrange the quadrature operators and the phase space variables
 into four dimensional vectors 
\mbox{$\hat{{\bf X}}=(\hat{x}_1,\hat{p}_1,\hat{x}_2,\hat{p}_2)$}
 and \mbox{${\bf X}=(x_1, p_1, x_2, p_2)$}, and define a \textit{quadrature variance matrix}~\cite{Sim3}:
\begin{eqnarray}
V_{\alpha\beta}=\langle \{ \Delta \hat{X}_{\alpha},\Delta \hat{X}_{\beta} \} \rangle
=Tr (\{ \Delta \hat{X}_{\alpha},\Delta \hat{X}_{\beta}\} \hat{\rho})\nonumber\\
\label{variance}
=\int d^4 X
\Delta X_{\alpha}\Delta X_{\beta}
W(\bf{X}),
\end{eqnarray}
where $\{ \Delta \hat{X}_{\alpha},\Delta \hat{X}_{\beta} \}=(\Delta \hat{X}_{\alpha}\Delta \hat{X}_{\beta}+\Delta
\hat{X}_{\beta}\Delta \hat{X}_{\alpha})/2$. Of course $\Delta \hat{X}_{\alpha}= \hat{X}_{\alpha} -
\langle \hat{X}_{\alpha} \rangle$ and $\langle \hat{X}_{\alpha} \rangle = Tr (\hat{X}_{\alpha}
\hat{\rho})$.
Similarly, $\Delta X_{\alpha} = X_{\alpha} -
\langle X_{\alpha} \rangle$, where $\langle X_{\alpha} \rangle$ is calculated as an average of $X_{\alpha}$ with respect to the Wigner
distribution $W({\bf X})$. Let us stress that $\langle X_{\alpha} \rangle$ equals to $\langle
\hat{X}_{\alpha} \rangle$.

A (zero-mean) Gaussian state is described by the Wigner distribution in a form:
\begin{equation}
\label{wigner} W({\bf X}) = (2\pi)^{-2}(\det{\bf V})^{-1/2} \exp (-\frac{1}{2} {\bf
X} {\bf V}^{-1} {\bf X}^T),
\end{equation}
where ${\bf V}$ is the variance matrix defined in eq.~(\ref{variance}).

To verify whether our system is in an entangled state we use the generalization of the Peres-Horodecki criterion~\cite{Per,Hor}
for continuous variables, derived by
 Simon~\cite{Sim}. In terms of the quadrature variance matrix, eq.~(\ref{variance}), rewritten as follows:
\begin{equation}
{\bf V}=\left[\begin{array}{cc}
{\bf A} & {\bf C}\\
{\bf C}^T & {\bf B}
\end{array}\right],
\end{equation}
 it takes a form of an inequality~\cite{Sim}:
\begin{eqnarray}
\label{simoncrit}&&det{\bf A}det{\bf B}+(\frac{1}{4}-|det{\bf C}|)^2\\
&&-Tr[ {\bf A}{\bf J}{\bf C}{\bf J}{\bf B}{\bf J}{\bf C}^T{\bf
J}]-\frac{1}{4} (det{\bf A}+det{\bf B})\ge 0,\nonumber
\end{eqnarray}
where matrix ${\bf J}$ is: 
\begin{equation}
{\bf J}=\left[\begin{array}{cc}
0 & 1\\
-1 & 0
\end{array}
\right].
\end{equation}
 Inequality~(\ref{simoncrit}) is a \textit{necessary} condition on the variance
matrix of a separable bipartite state~\cite{Sim}.
 Moreover, Simon~\cite{Sim} has shown that this is also a \textit{sufficient} condition for separability
 for all bipartite Gaussian states composed of two modes, each mode held by each party~\cite{Eisert}. Thus for Gaussian states considered in this paper, when the inequality~(\ref{simoncrit}) is violated the state is
 entangled, otherwise it is separable.

Let us note, that there is possible an equivalent approach to the separability properties of quantum two-party
Gaussian states in the framework of the operator formalism for the density operator, as it has
been shown by Englert and W\'odkiewicz~\cite{englert}.

To compute the degree of entanglement we use the {\it logarithmic negativity}~\cite{vidal}:
\begin{equation}\label{miara}
E(\rho)=log_2||\hat{\rho}^{T_A}||_1,
\end{equation}
where $||\hat{\rho}^{T_A}||_1$ denotes the trace norm of $\hat{\rho}^{T_A}$, and
$\hat{\rho}^{T_A}$ is a
partial transpose of the bipartite state $\hat{\rho}$ with respect to the subsystem $A$. This measure can be regarded as a
quantitative version of the Peres-Horodecki criterion~\cite{Per,Hor}, because it quantifies
the degree to which $\hat{\rho}^{T_A}$ fails to be positive~\cite{vidal}.

For states described by the Wigner function of the form~(\ref{wigner}), computing of the
logarithmic negativity~(\ref{miara}) simplifies to process known as sympletic
diagonalization of the variance matrix~(\ref{variance}), as has been shown in~\cite{vidal},
 i.e. if $(\lambda_1,\lambda_2)$ is the
sympletic spectrum of the variance matrix~(\ref{variance}), then the logarithmic negativity equals:
\begin{equation}
E(\rho)= F(\lambda_1)+F(\lambda_2),
\end{equation}
where $F(\lambda_i)$ ($i=1,2$) is:
\begin{equation}\label{f}
F(\lambda_i)=\left\{ \begin{array}{cc}
 0 & {\rm for}\  2\lambda_i\ge1,\\
-log_2(2\lambda_i) & {\rm for}\  2\lambda_i<1.
\end{array}\right.
\end{equation}

Evolution of two modes of the electromagnetic field coupled to the common thermal reservoir might be described by the
diffusion equation, i.e. the Fokker-Planck equation. Namely, for the interaction Hamiltonian of the
form~(\ref{jedenham}) it reads (in the interaction picture):
\begin{eqnarray}
\label{F-P}\partial_t W({\bf X}, t)& = & \sum_{j,k=1,2}  [ \frac{\gamma}{2} (\partial_{x_j}x_k + \partial_{p_j}p_k)\nonumber\\
&+&\frac{2\bar{N}+1}{4} (\partial_{x_j,x_k}^2 +\partial_{p_j,p_k}^2) ] W({\bf X},t).\nonumber\\
\end{eqnarray}
where $\gamma$ is a coupling constant and $\bar{N}$ is a mean thermal photon number. Here it is
worthwhile to notice that in the case, when one couples each mode to its own reservoir, i.e. the
interaction Hamiltonian is of the form~(\ref{dwaham}), the relevant Fokker-Planck equation is
obtained from eq.~(\ref{F-P}) by substitution $j=k$. 

\section{Results}
The process of the parametric down conversion in the non-linear
 crystals is used as a source
  of entangled pairs in many experiments~\cite{wu,Fur}.
   In such process two-mode squeezed vacuum states are produced. Thus as an initial state of our
 system we consider the  two-mode squeezed vacuum, which is described by the
 Wigner function of the form~(\ref{wigner}) with
the variance matrix as follows:
\begin{equation}
\label{var} {\bf V} =\frac{1}{2}\left[\begin{array}{cccc}
 n_1 & 0 & c_1 & 0\\
 0 & n_2 & 0 & c_2\\
 c_1 & 0 & n_1 & 0\\
 0 & c_2 & 0 & n_2 \end{array}\right],
\end{equation}
where $n_i$ and $c_i$ are:
\begin{eqnarray}\label{poczatek}
&n_i = \cosh (2r),\nonumber\\
\label{nici}&c_2=-c_1=\sinh (2r),
\end{eqnarray}
where $r$ is a squeezing parameter (for $r\ne0$ this state is entangled). Without loss of generality we
have assumed that the squeezing parameter, $r$, is real.

While our system evolves, the general form of the variance matrix, eq.~(\ref{var}), remains constant, but elements
$n_i$ and $c_i$ are changing. Thus we may rewrite Simon criterion, eq.~(\ref{simoncrit}), in a form:
\begin{equation}\label{nierow}
\frac{1}{16} [(n_1-|c_1|)(n_2-|c_2|)-1][(n_1+|c_1|)(n_2+|c_2|)-1]\ge0.
\end{equation}

Simon criterion, eq.~(\ref{simoncrit}), is a yes-no test for entanglement and thus only the sign of the left
hand side of the inequality~(\ref{nierow}) is important. Therefore we omit an expression, which does not change
the sign of the inequality~(\ref{nierow}). Namely, because $|c_i|$ is always positive, unless it is equal to
zero and $n_1n_2$ is always greater than one, what is a consequence of the uncertainty principle, it is more convenient to check the sign of the following inequality, instead of using
the inequality~(\ref{nierow}),~\cite{kim}:
\begin{eqnarray}
\label{kimcrit}(n_1-|c_1|)(n_2-|c_2|)-1\ge 0
\end{eqnarray}
When it is
satisfied then state is separable, otherwise it is entangled.

 Solving eq.~(\ref{F-P}) with the initial variance matrix in the form of eq.~(\ref{var}) with $n_i$
 and $c_i$ given in eq.~(\ref{nici}) leads to time-dependent form of $n_i$ and $c_i$:
\begin{eqnarray}
\label{elementy}
& n_1=\frac{1}{2}[2\cosh (2r) + (N-e^{-2r})\tau ],\nonumber\\
& n_2=\frac{1}{2}[2\cosh (2r) + (N-e^{2r})\tau ],\\
& c_1=\frac{1}{2}[(N-e^{-2r})\tau -2\sinh (2r)],\nonumber\\
& c_2=\frac{1}{2}[(N-e^{2r})\tau +2\sinh (2r)],\nonumber
\end{eqnarray}
where $\tau =1-e^{-2\gamma t}$, $N=2\bar{N}+1$ and $\gamma$ denotes coupling. Then substituting~(\ref{elementy}) into the inequality~(\ref{kimcrit}) we find that if the
initial state is sufficiently squeezed, i.e.:
\begin{equation}
\label{mojwaruneknascisniecie}|r|\ge\frac{1}{2}\ln (2\bar{N}+1),
\end{equation}
it will remain entangled forever in spite of interaction. 
 Otherwise, the state
will disentangle after time:
\begin{equation}
\label{mojczas1}t=\frac{1}{2\gamma}\ln\left(\frac{2\bar{N}+1-e^{-2|r|}}{2\bar{N}+1-e^{2|r|}}\right).
\end{equation}

Thus basing on the condition~(\ref{mojwaruneknascisniecie}) it is possible to choose such initial state of the
two-mode squeezed vacuum, for which the entanglement
will survive the interaction with the common reservoir.

Inequality~(\ref{mojwaruneknascisniecie}) seems to have the same form as a condition for
non-classicality of a squeezed thermal state~\cite{kim3}. In the latter case, however, $\bar{N}$ is a mean thermal photon
number and $r$ is a squeezing parameter, both corresponding to the thermal squeezed state. Here
$\bar{N}$ corresponds to a number of thermal photon in the environment and $r$ is a squeezing parameter of a
state embedded in this environment.

For a comparison, if one takes into consideration the coupling to two independent thermal reservoirs
, it will
lead to the same general form of the quadrature variance matrix, eq.~(\ref{var}), and the separability
condition,
eq.~(\ref{kimcrit}), but coefficients $n_i$ and $c_i$ would be different, namely~\cite{kim2,kim,duan}:
\begin{eqnarray}\label{elementykim}
& n_i=\cosh (2r) e^{-\gamma t} + N(1-e^{-\gamma t}),\\
& c_2=-c_1=\sinh (2r) e^{-\gamma t},\nonumber
\end{eqnarray}
where, as before, $r$ is a squeezing parameter, $N=2\bar{N}+1$ and $\gamma$ denotes coupling. 
In such a case, every state will disentangle after the time:
\begin{equation}
\label{czasinni}t=\frac{1}{\gamma}\ln \left(1+\frac{1-e^{-2|r|}}{2\bar{N}}\right),
\end{equation}
unless it would interact with vacuum,
as it has been shown by Duan \textit{et al.}~\cite{duan}, Scheel \textit{et
al.}~\cite{welsch} and Paris~\cite{paris}. Such a situation corresponds to the limit $\bar{N}\rightarrow0$, what implies
$t\rightarrow\infty$.

For the variance matrix of the form~(\ref{var}) it is easy to find the sympletic spectrum and the
logarithmic negativity. The sympletic spectrum takes form:
\begin{eqnarray}\label{lambda}
\lambda_1=\frac{1}{2}\sqrt{(n_1-c_1)(n_2+c_2)},\nonumber\\
\lambda_2=\frac{1}{2}\sqrt{(n_1+c_1)(n_2-c_2)}.
\end{eqnarray}
Using above mentioned sympletic spectrum, eq.~(\ref{lambda}),
 and conditions for function $F(\lambda_i)$, eq.~(\ref{f}),
  together with eq.~(\ref{elementy}) it is possible to derive
inequality~(\ref{mojwaruneknascisniecie}) and expression for the disentanglement time in the
one reservoir case,
eq.~(\ref{mojczas1}). Analogously it is possible to derive formula for the disentanglement time
in the case of the interaction with two reservoirs, eq.~(\ref{czasinni}).

Especially, sympletic spectrum, eq.~(\ref{lambda}), holds for initial two-mode
squeezed vacuum state with $n_i$ and $c_i$ given by eq.~(\ref{poczatek}), and
leads to conclusion that the logarithmic negativity for such state is simply
propotional to the absolute value of the squeezing parameter, $r$:
\begin{equation}
E(\rho)=\frac{2}{\ln {2}} |r|.
\end{equation}

Moreover, having time-dependent form of $n_i$ and $c_i$, eq.~(\ref{elementy}), allows us
to write down the analytic expression for the asymptotic value of the logarithmic negativity at
$t\rightarrow\infty$:
\begin{equation}
E(\rho)=\frac{1}{\ln 2}|r|-\frac{1}{2}\log_2 (2\bar{N}+1),
\end{equation}
which is valid for initial states fulfilling
condition~(\ref{mojwaruneknascisniecie}). In case, when
$|r|=\frac{1}{2}\ln(2\bar{N}+1)$, $\lim_{t\to\infty}E(\rho)=0$ (compare with
event of the interaction of two-mode squeezed state with two separate reservoirs in a
vacuum state).
 
\begin{figure}[t]
\includegraphics[width=0.4\textwidth,height=0.3\textwidth]{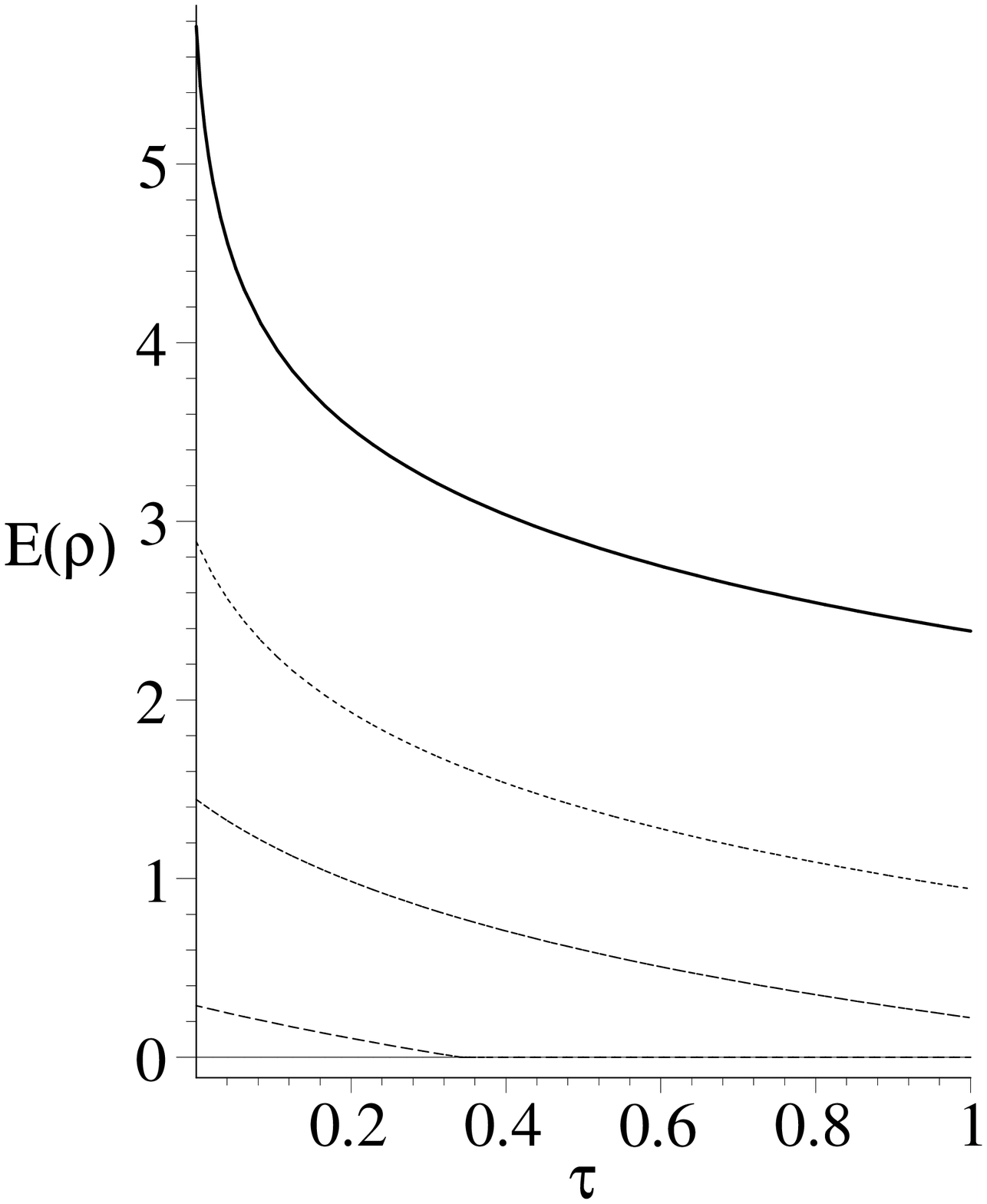}
\caption{Logarithmic negativity, $E(\rho)$, for the two-mode squeezed state as function of a
dimensionless time, $\tau=1-exp(-2\gamma t)$, which is $0$ for $t=0$ and $1$ for $t=\infty$, in the common reservoir
model, for $\bar{N}=0.5$ and $r=(0,0.1,0.5,1,2)$ (from the bottom to the top).\label{onereservoir1}}
\end{figure}

\begin{figure}[t]
\includegraphics[width=0.4\textwidth,height=0.3\textwidth]{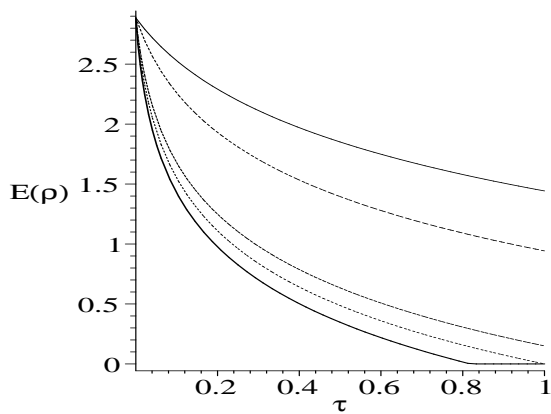}
\caption{Logarithmic negativity, $E(\rho)$, for the two-mode squeezed state as function of a
dimensionless time, $\tau=1-exp(-2\gamma t)$, which is $0$ for $t=0$ and $1$ for $t=\infty$, in the common reservoir
model, for $r=1$ and $N=(1,2,6,exp(2),9)$ (from the top to the bottom), $N=2\bar{N}+1$.\label{onereservoir2}}
\end{figure}

Figs.~\ref{onereservoir1}
and~\ref{onereservoir2} show changes of the logarithmic negativity, $E(\rho)$,
 during the interaction with the common reservoir for different initial values of
the squeezing parameter, $r$, and the mean thermal photon number, $\bar{N}$.
 Logarithmic negativity, $E(\rho)$, is plotted as a
function of the dimensionless rescaled time $\tau=1-\exp(-2\gamma t)$ which equals to zero
 for $t=0$ and to one for $t=\infty$. Note that the coupling constant, $\gamma$, changes
  only the time scale and has no effect
on the logarithmic negativity itself. From Fig.~\ref{onereservoir1}
it is easy
to notice that states satisfying condition~(\ref{mojwaruneknascisniecie}) (i.e. for
$r=0.5$, $r=1$ and $r=2$) remain entangled forever
in spite of interaction.

\begin{figure}[t]
\includegraphics[width=0.4\textwidth,height=0.3\textwidth]{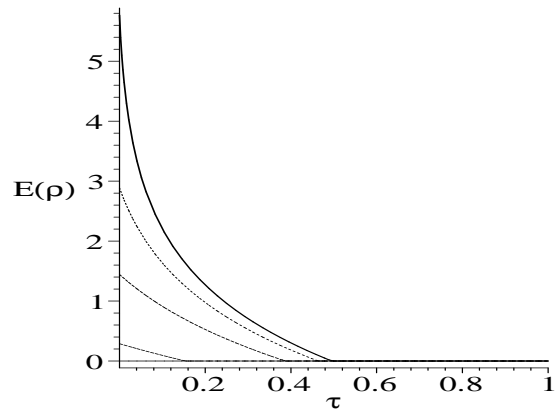}
\caption{Logarithmic negativity, $E(\rho)$, for the two-mode squeezed state as function of a
dimensionless time, $\tau=1-exp(-\gamma t)$, which is $0$ for $t=0$ and $1$ for
$t=\infty$, in the two reservoirs
model, for $\bar{N}=0.5$ and $r=(0,0.1,0.5,1,2)$ (from the bottom to the top).\label{tworeservoirs1}}
\end{figure}

\begin{figure}[t]
\includegraphics[width=0.4\textwidth,height=0.3\textwidth]{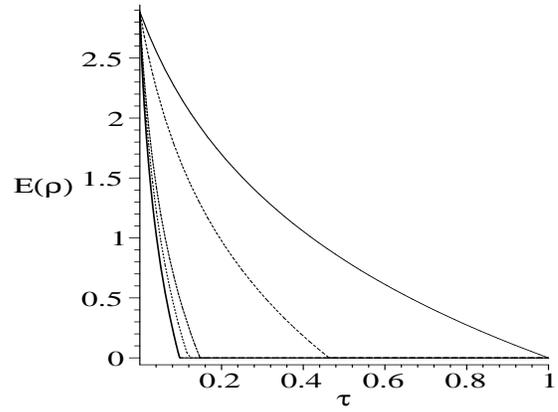}
\caption{Logarithmic negativity, $E(\rho)$, for the two-mode squeezed state as function of a
dimensionless time, $\tau=1-exp(-\gamma t)$, which is $0$ for $t=0$ and $1$ for
$t=\infty$, in the two reservoirs
model, for $r=1$ and $N=(1,2,6,exp(2),9)$ (from the top to the bottom), $N=2\bar{N}+1$.\label{tworeservoirs2}}
\end{figure}

Figs.~\ref{tworeservoirs1}
and~\ref{tworeservoirs2} show changes of the logarithmic negativity, $E(\rho)$, during
 the interaction with two separated reservoirs for
different values of the squeezing parameter, $r$, and the mean thermal photon number,
$\bar{N}$. Again the logarithmic negativity, $E(\rho)$,
 is plotted as a function of the dimensionless
time $\tau=1-\exp(-\gamma t)$ which equals to zero for $t=0$ and to one for $t=\infty$.
 From Fig.~\ref{tworeservoirs2} it is easy to notice that only in the case of interaction
  with vacuum, namely $N=1$ states remain entangled. 

 Let us stress that Figs.~\ref{onereservoir1} and~\ref{tworeservoirs1} (similarly
 Figs.~\ref{onereservoir2} and~\ref{tworeservoirs2}) are plotted for the same initial values of
the squeezing parameter, $r$ and the mean thermal photon number, $\bar{N}$, but for different models of interaction.

It is worth noticing that in a case, when a pure two-mode squeezed vacuum state interacts with
 a common reservoir initially in vacuum state ($\bar{N}=0$) the resulting state, after infinitely
long interaction, is a pure, but it is not a vacuum. Final state is squeezed, and thus entangled.

That is the case because of properties of the interaction Hamiltonian~(\ref{jedenham}), which show up
after non-local transformation of variables, such as:
\begin{equation}\label{zmiana}
\left(\begin{array}{c}
x_1\\
p_1\\
x_2\\
p_2\end{array}\right)
\to \left(\begin{array}{c}
x_S\\
p_S\\
x_D\\
p_D\end{array}\right)
=\left(\begin{array}{c}
x_1+x_2\\
p_1+p_2\\
x_2-x_1\\
p_2-p_1\end{array}\right).
\end{equation}
Were new variables are just sums and differences of the old ones.

\begin{figure}[t]
\includegraphics[width=0.4\textwidth,height=0.3\textwidth]{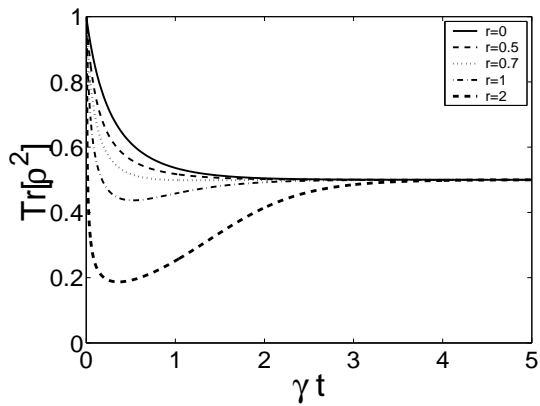}
\caption{Purity, $Tr[\rho^2]$, for the two-mode squeezed state as function of $\gamma t$, in the common reservoir
model, for $\bar{N}=0.5$ and different initial squeezing.\label{onerespur1}}
\end{figure}

\begin{figure}[t]
\includegraphics[width=0.4\textwidth,height=0.3\textwidth]{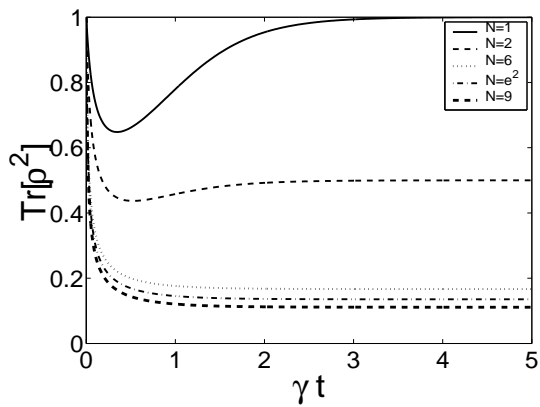}
\caption{Purity, $Tr[\rho^2]$, for the two-mode squeezed state as function of $\gamma t$, in the common reservoir
model, for $r=1$ and differen mean thermal photon number, $N=2\bar{N}+1$.\label{onerespur2}}
\end{figure}

 Our initial state is pure and rewriting it in new basis does not change its purity,
while the interaction Hamiltonian~(\ref{jedenham}) appears to couple to the reservoir only mode $S$ described by sums,
$(x_S,p_S)$, leaving mode $D$ described by differences, $(x_D,p_D)$, unaffected~\cite{ekert}.
Thus, only mode $S$ interacts with thermal reservoir and mode $D$ undergoes free evolution. As a
consequence, mode $S$ first experiences decoherence and then dissipation to the state of the reservoir
(see Figs.~\ref{onerespur1}
and~\ref{onerespur2} - decoherence for $\gamma t = 0-0.5$ and dissipation for $\gamma t >
0.5$).
Therefore, if the vacuum is the environment of interest, mode $S$ becomes vacuum after some time.
Clearly both modes are pure, mode $S$ because it is vacuum and mode $D$ because it had evolved freely.
Hence the final state is also pure, but it is still squeezed, because some squeezing ''survived'' in mode
$D$, and thus it is still entangled. If the reservoir initially is not in vacuum, the final state will
be mixed. Here mode $D$ enables to span some kind of a decoherence-free subspace~\cite{dfree}.

Of course in the case of interaction with two independent reservoirs, the interaction
Hamiltonian~(\ref{dwaham}) do not exhibit such property under change of variables~(\ref{zmiana}).

For the sake of clarity we plot the purity, $Tr[\rho^2]$, of the state of interest as a function of time for different
initial values of
the squeezing parameter, $r$, and the mean thermal photon number, $\bar{N}$ (see Figs.~\ref{onerespur1}
and~\ref{onerespur2}).
For a Gaussian state described by the Wigner function with the variance matrix of the form~(\ref{var}) purity equals:
\begin{equation}
Tr[\rho^2]=\frac{1}{\sqrt{(n_1^2-c_1^2)(n_2^2-c_2^2)}}.
\end{equation}

\section{Summary}
Summarizing, for the coupling to the common reservoir, we noticed that whether the state is
entangled or not depends mostly on the initial degree of squeezing. If the state is initially
sufficiently squeezed, i.e. according to condition~(\ref{mojwaruneknascisniecie}), it will
always remain entangled, independently of the strength of the coupling to the
environment. Otherwise, the state initially
entangled will become separable after the time given in the eq.~(\ref{mojczas1}). It is clearly
different result
from that obtained for the model in which each of two entangled modes interacts with its own reservoir. In
such a case only the interaction with vacuum fluctuations does not lead to the disentanglement ~\cite{duan}. 
Moreover, we have shown that in a case of interaction with a common reservoir being in a vacuum state,
 the state of interest, after infinitely long interaction is not only still entangled, but also it is pure. 

\section{Acknowledgments}
Discussions with Prot Pako\'nski, Jacek Dziarmaga, Artur Ekert, Zbyszek Karkuszewski, Kuba Zakrzewski and Karol
\.Zyczkowski are acknowledged. The work was supported by the Polish Ministry of Scientific Research
Grant Quantum Information and Quantum Engineering PBZ-MIN-008/P03/2003 and by the Stefan
Batory Foundation (Warszawa, Poland), through the scholarship at Oxford University.


\begin{thebibliography}{99}
\bibitem{Bar} A. Barenco, D. Deutsch, A. Ekert and R. Jozsa, Phys. Rev. Lett. {\bf 74, }4083 (1995).
\bibitem{Ben} C. H. Bennett, G. Brassard, C. Crepeau, R. Jozsa, A. Peres and W. K. Wootters, Phys. Rev. Lett. {\bf 70, }1895 (1993).
\bibitem{braun} S. L. Braunstein and H. J. Kimble, Phys. Rev. Lett. {\bf 80, }869(1998).
\bibitem{wu} B. L. Schumaker and C. M. Caves, Phys. Rev. A {\bf 31, }3093 (1985); L. A. Wu, H. J.
Kimble, J. L. Hall, and H. Wu, Phys. Rev. Lett. {\bf 57, }2520 (1986); S. M. Barnett and P. L.
Knight, J. Mod. Opt. {\bf 34, }841 (1987).
\bibitem{Fur} A. Furusawa et al., Science {\bf 282, }706 (1998).
\bibitem{kim2} H. Jeong, J. Lee, and M. S. Kim, Phys. Rev. A {\bf 61, }052101 (2000); J. Lee, M. S. Kim, and
H. Jeong, Phys. Rev. A {\bf 62, }032305 (2000).
\bibitem{kim} M. S. Kim and J. Lee, e-print quant-ph/0203151.
\bibitem{welsch}  S. Scheel, T. Opatrn\'y and D.-G. Welsch, e-print quant-ph/0006026; S. Scheel and
D.-G. Welsch, Phys. Rev. A {\bf 64, }063811 (2001).
\bibitem{hiroshima} T. Hiroshima, Phys. Rev. A {\bf 63, }022305 (2001).
\bibitem{paris} M. G. A. Paris, J. Opt. B {\bf 4, }442 (2002).
\bibitem{Bowen} W. P. Bowen, R. Schnabel, P. K. Lam and T. C. Ralph, e-print quant-ph/0209001.
\bibitem{Wigner} E. P. Wigner, Phys. Rev. {\bf 40, }749 (1932).
\bibitem{Sim3} R. Simon, E. C. G. Sudarshan and N. Mukunda, Phys. Rev. A {\bf
36, }3868 (1987); R. Simon, N. Mukunda and B. Dutta, Phys. Rev. A {\bf 49, }1567 (1994).
\bibitem{Per} A. Peres, Phys. Rev. Lett.  {\bf 77, }1413 (1996).
\bibitem{Hor} M. Horodecki, P. Horodecki and R. Horodecki, Phys. Lett. A {\bf 223, }1 (1996); P. Horodecki, Phys. Lett. A {\bf 232, }333 (1997).
\bibitem{Sim} R. Simon, Phys. Rev. Lett.  {\bf 84, }2726 (2000).
\bibitem{Eisert} In fact, for a bipartite Gaussian state such that one party
holds one mode and the other party holds $N$ modes positivity of the partial
transpose is sill necessary and sufficient criterion for separability. See J.
Eisert and M. B. Plenio, e-print quant-ph/0312071; G. Giedke, L.-M. Duan, J.
I. Cirac, and P. Zoller, Quant. Inf. Comp. {\bf 1, }79 (2001); R. F. Werner
and M. M. Wolf, Phys. Rev. Lett. {\bf 86, }3658 (2001); G. Giedke, B. Kraus,
M. Lewenstein, and J. I. Cirac, Phys. Rev. Lett. {\bf 87, }167904 (2001).
\bibitem{englert} B. G. Englert and K. W\'odkiewicz, Phys. Rev. A {\bf 65, }054303 (2002).
\bibitem{vidal} G. Vidal and R. F. Werner, Phys. Rev. A {\bf 65, }032314 (2002).
\bibitem{kim3} M. S. Kim, F. A. M. de Oliveira, and P. L. Knight, Phys. Rev. A {\bf 40, }2494 (1989).
\bibitem{duan} L. M. Duan, G. Giedke, J. I. Cirac and P. Zoller, Phys. Rev. Lett. {\bf 84,
}2722 (2000).
\bibitem{ekert} A. K. Ekert, {\it Correlation in Quantum Optics, }PhD Thesis University of Oxford, OUALP-92-1, submitted in 1991.
\bibitem{dfree} D. A. Lidar and K. B. Whaley, {\it Decoherence-free Subspaces and Subsystems,}
 in "Irreversible Quantum Dynamics",
 F. Benatti and R. Floreanini (Eds.), pp. 83-120 (Springer Lecture Notes in Physics vol. 622, Berlin, 2003).



\end{thebibliography}
\end{document}